\documentclass[12pt]{article}
\usepackage{amssymb}
\usepackage{amsmath}
\usepackage{epsfig}
\usepackage{float}
\newcommand{\be}{\begin{equation}}
\newcommand{\ee}{\end{equation}}
\newcommand{\bea}{\begin{eqnarray}}
\newcommand{\eea}{\end{eqnarray}}
\begin{document}

\begin{center}
{\bf NEUTRINO OSCILLATIONS AND TIME-ENERGY UNCERTAINTY RELATION }

\end{center}

\begin{center}
S. M. Bilenky
\end{center}

\begin{center}
{\em  Joint Institute
for Nuclear Research, Dubna, R-141980, Russia, and\\
SISSA,via Beirut 2-4, I-34014 Trieste, Italy.}
\end{center}

\abstract{Neutrino oscillations is a phenomenon which is characterized by a finite oscillation time 
(length). For such phenomena time-energy uncertainty relation is valid.
This means that  energy uncertainty is needed for oscillations to occur. 
We consider  neutrino
oscillations from this point of view. We demonstrate that for mixed neutrino states, which describe 
flavor neutrinos $\nu_{e}$, $\nu_{\mu}$ and $\nu_{\tau}$, translational invariance does not take place.} 

\section{Introduction}

Evidence of neutrino oscillations
obtained in the atmospheric Super-Kamiokande \cite{SK}, solar SNO \cite{SNO}, 
reactor KamLAND \cite{Kamland} and other neutrino experiments \cite{K2K,Cl,Gallex,Sage,SKsol}
 is one of the most important recent discovery in particle physics. 

All existing neutrino oscillation data (with the exception of the LSND data)\footnote{Indication in favor of $\bar\nu_{\mu}\leftrightarrows\bar\nu_{e} $ oscillation obtained several years ago in 
the accelerator short-baseline  LSND experiment \cite{LSND} will be checked by the running at the Fermilab 
MiniBooNE experiment \cite{Miniboone}.} are described by the three-neutrino mixing.
The character of neutrino oscillations, observed in present-day experiments, is determined by 
two inequalities:
\begin{enumerate}
\item 
$\Delta m_{12}^{2}\ll \Delta m_{23}^{2}.$ 

\item
$\sin^{2}\,\theta_{13}\ll 1.$
\end{enumerate}
Here 
$$\Delta m^{2}_{ik}= m^{2}_{k}- m^{2}_{i}$$ and $m_{i}$ is neutrino mass.
The first inequality is based on the analysis of all existing data. For best-fit values we have
\be
\Delta m_{12}^{2}\simeq 3.3\cdot 10^{-2}~ \Delta m_{23}^{2}. 
\label{1}
\ee
The second inequality follows from 
the results of CHOOZ reactor experiment \cite{Chooz}. From the exclusion plot obtained from the analysis of the data of this experiment it can be  found 
\be
\sin^{2}\,\theta_{13}\leq 5\cdot 10^{-2}.
\label{2}
\ee
In the case of atmospheric and accelerator long-baseline experiments (SK \cite{SK}, K2K \cite{K2K}, 
MINOS \cite{Minos}, OPERA \cite{Opera})
in the leading approximation 
we can neglect contribution of $\Delta m_{12}^{2}$ and $\sin^{2}\,\theta_{13}$ to the probability of neutrino transitions in 
vacuum. In this approximation neutrino oscillations are  two-neutrino  
$\nu_\mu \leftrightarrows \nu_\tau$ ($\bar\nu_\mu \leftrightarrows \bar\nu_\tau$) oscillations (see \cite{BGG}).
From the analysis of the data of the atmospheric Super-Kamiokande experiment 
the following 90 \% CL ranges were obtained \cite{SK}
\be 
1.5\cdot 10^{-3}\leq \Delta m^{2}_{23} \leq 3.4\cdot
10^{-3}\rm{eV}^{2};~~ \sin^{2}2 \theta_{23}> 0.92.\label{3}
\ee
In the case of  the reactor KamLAND experiment and 
solar experiments
in the leading approximation  we can neglect  contribution of 
$\sin^{2}\,\theta_{13}$ to the $\nu_{e}$ survival
probability in vacuum and ,correspondingly, in matter.
Neutrino oscillations in this approximation are 
$\nu_e \leftrightarrows \nu_{\mu,\tau}$ ($\bar\nu_e \leftrightarrows \bar\nu_{\mu,\tau}$).
Transition probability in vacuum and, correspondingly, in matter  has two-neutrino form and depend on parameters 
$\Delta m^{2}_{12}$  and $\tan^{2} \theta_{12}$. 
From analysis of all solar and KamLAND data 
(assuming CPT invariance)
it was obtained \cite{SNO}
\be 
\Delta m^{2}_{12} = 8.0^{+0.6}_{-0.4}~10^{-5}~\rm{eV}^{2};~~~
\tan^{2} \theta_{12}= 0.45^{+0.09}_{-0.07}.\label{4}
\ee
In spite of great progress in the investigation of neutrino oscillations, there are many open problems in physics of massive and mixed neutrinos.
\begin{enumerate}
\item The nature of neutrinos with definite masses (Majorana or Dirac?) is unknown.
To reveal Majorana nature of neutrinos necessary to study neutrinoless double 
$\beta$-decay of some even-even nuclei.
Several new experiments are in preparation (see \cite{bbfuture}).

\item For absolute values of neutrino masses only upper bounds are known.  From Mainz \cite{Mainz} 
 and Troitsk \cite{Troitsk}
tritium experiments it was found 
\be
m_{\beta}\leq 2.3~ \rm{eV},
\label{5}
\ee
were $m_{\beta}\simeq m_{0}$, $m_{0}$ is common neutrino mass.
From cosmological data for the sum of neutrino masses 
upper bounds in the range
\be
\sum_{i} m_{i}\leq (0.4-1.7)~ \rm{eV}
\label{6}
\ee
were inferred (see \cite{Tegmark}).
With new experiments further progress is expected.
\item 
For the parameter $\sin^{2}\,\theta_{13}$ only upper bound (\ref{2}) is known.
If the value of this parameter is not very small such effects of the three-neutrino mixing as 
CP-violation can be studied.
New reactor and accelerator experiments in which the value of the parameter $\sin^{2}\,\theta_{13}$
will be measured (or CHOOZ bound will be improved) are in preparation \cite{2Chooz}.

\item The character of the neutrino mass spectrum 
is unknown. Information about character of neutrino mass spectrum can be obtained from future 
accelerator experiments \cite{T2K} and from $0\nu\beta\beta$ experiments.

\end{enumerate}

In spite the existence of neutrino oscillations was established the 
basics of this new phenomenon is still a subject of active discussions 
(see review \cite{Carlo} and many references therein). We will 
consider here neutrino oscillations from the point of view of time-energy uncertainty relation
which take place  for any quantum phenomena with a characteristic time 
during which the state of the system is significantly changed.

\section{Flavor neutrino states}
From the point of view of the field theory neutrino oscillations are based on the fact 
that flavor fields  $\nu_{lL}(x)$ in leptonic charged and neutral currents
\be 
j^{\mathrm{CC}}_{\alpha}(x) =2 \sum_{l=e,\mu,\tau} \bar \nu_{lL}(x)
\gamma_{\alpha}l_{L}(x);~~ j ^{\mathrm{NC}}_{\alpha}(x)
=\sum_{l=e,\mu,\tau} \bar \nu_{lL}(x)\gamma_{\alpha}\nu_{lL}(x)
\label{7}
\ee
are {\em mixed fields}
\be
\nu_{l L}(x) = \sum^{3}_{i=1} U_{{l}i} ~ \nu_{iL}(x). \label{8}
\ee
Here $\nu_{i}(x)$ is (Majorana or Dirac) field of neutrino with mass $m_{i}$ and 
$U$ is the unitary PMNS \cite{BP,MNS}
mixing matrix. The neutrino masses, mixing angles and CP phase are parameters which are 
determined by a
 generation mechanism (still unknown).

Neutrinos are produced in CC weak decays and reactions. Let us consider, for example, production of a lepton $l^{+}$ and neutrino in a decay (see \cite{BilG})
\be
a \to b + l^{+}+ \nu_{i},~~~(l=e, \mu,\tau)
\label{9}
\ee
where $a$ and $b$ are some hadrons. Normalized final neutrino state is given by
\be
|\nu_{f}\rangle =\frac{1}{\sum_{i}|\langle \nu_i \,l^{+}\,b\,| S |\,a\rangle|^{2}}  \sum_{i}|\nu_i\rangle  ~ \langle \nu_i \,l^{+}\,b\,| S |\,a\rangle,
\label{10}
\ee
where $\langle \nu_i \,l^{+}\,b\,| S |\,a\rangle$ is the matrix element of the process (\ref{9})
and $|\nu_i\rangle $ is the state of left-handed neutrino with mass $m_{i}$, momentum 
$\vec{p}$ and energy  $E_{i}= \sqrt{p^{2}+ m^{2}_{i}}\simeq p +\frac{m^{2}_{i}}{2p}$. 

Neutrino energies $E$ in neutrino oscillation experiments are much larger than neutrino masses
(in solar and reactor experiments $E\gtrsim 1$ MeV, in atmospheric and accelerator long-baseline experiments  $E\gtrsim 1$ GeV etc). 
Due to Heisenberg uncertainty relation it is 
impossible to distinguish 
production of different neutrinos in neutrino production processes.
Let us consider, for example, the decay $\pi^{+}\to \mu^{+} +\nu$ in the pion rest-frame. 
 Difference of momenta of 
neutrinos $\nu_{i}$ and $\nu_{k}$ is given by 
\be
\Delta p_{ik} = p_{k}-p_{i} \simeq -\frac{\Delta m^{2}_{ik}}{2 E}(1-\frac{E}{m_{\pi}}).
\label{11}
\ee
Here 
\be
E=\frac{m^{2}_{\pi}-m^{2}_{\mu}}{2m_{\pi} }\simeq 29.8 ~\rm{MeV}.
\label{12}
\ee
We can not distinguish production of $\nu_{i}$ and $\nu_{k}$ if
the following inequality takes place
\be
|\Delta p_{ik}|\lesssim \Delta p_{\rm{QM}}\simeq \frac{1}{d}
\label{13}
\ee
Here $\Delta p_{\rm{QM}}$ is quantum-mechanical uncertainty of neutrino momentum, and $d$  
characterizes
a quantum-mechanical size of neutrino source. We can rewrite the condition  (\ref{13}) in the form 
\be
L_{ik} \gtrsim d, 
\label{14}
\ee
where 
\be
L_{ik} =\frac{1}{|\Delta p_{ik}|}\simeq \frac{2E}{|\Delta m^{2}_{ik}|}
\label{15}
\ee
For the largest atmospheric neutrino mass-squared difference
$\Delta m^{2}_{23}=2.4 \cdot 10^{-3} \rm{eV}^{2}$,
\be
L_{23}\simeq 5\cdot 10^{5}~\rm{cm}
\label{16}
\ee
and the condition  (\ref{14}) is obviously  fulfilled.

Let us return now back to Eq.(\ref{10}). Taking into account that production of neutrinos with different 
masses can not be resolved,  from (\ref{7}) and (\ref{8}) we find
\be
\langle\nu_{ i}\,l^{+}\,b\,| S |\,a\rangle \simeq U_{l i}^{*}\,~
\langle \nu_{l}\,l^{+}\,b\,| S |\,a\rangle_{SM},
\label{17}
\ee
where $\langle \nu_{l}\,l^{+}\,b\,| S |\,a\rangle_{SM}$ is the Standard Model matrix element of the decay  (\ref{9}). We have\footnote{Because 
$E^{2}\gg m_{i}^{2}$ negligible contribution of neutrino masses to (\ref{18}) 
can be neglected.
Neutrino masses can be revealed by the measurement of the high-energy part of $\beta$-spectrum
of tritium which correspond to the emission of neutrino with energy
 comparable with neutrino mass.
Effect of neutrino masses can be observed in these  experiments if $m^{2}_{i}\gg \Delta m ^{2}_{23}$
and neutrino mass spectrum is quasi degenerate. Future KATRIN \cite{Katrin} tritium experiment will be sensitive to $m_{0}\simeq 0.2$ eV.}
 \be
\langle \nu_{l}\,l^{+}\,b\,| S |\,a\rangle_{SM}=-i\,\frac{G_{F}}{\sqrt{2}}\,N\,2\,
\bar u_{L}(p)\,\gamma_{\alpha}\,u_{L}(-p')\,\langle b|\, J^{\alpha}(0)\,|a \rangle \,(2\pi)^{4}\,
\delta (P'-P).
\label{18}
\ee
Here $N$ is the product of the standard normalization factors, $p'$ is the momentum of 
$l^{+}$, $p$ is neutrino momentum, $P$ and  $P'$ are total 
initial and final momenta and $J^{\alpha}$ is hadronic charged current.

From  (\ref{10}) and (\ref{17}) for the normalized neutrino left-handed state we find
\be
|\nu_{l}\rangle
= \sum_{i=1}^{3} U_{l i}^* \,~ |\nu_i\rangle.
\label{19}
\ee
Analogously, together with $l^{-}$ in CC processes right-handed antineutrino $\bar\nu_{l}$ 
are produced. 
The state of  $\bar\nu_{l}$ is given by
\be
|\bar\nu_{l}\rangle
= \sum_{i=1}^{3} U_{l i} \,~ |\nu_i\rangle,
\label{20}
\ee
where $|\nu_i \rangle$ is the state of right-handed neutrino (antineutrino in Dirac case) with mass 
$m_{i}$, momentum $\vec{p}$ and energy $E_{i}\simeq p +\frac{m^{2}_{i}}{2p}$. 

Thus, in CC neutrino production 
processes together with leptons  $l^{+}$ ($l^{-}$ ) flavor left-handed neutrinos $\nu_{l}$ 
(right-handed antineutrinos $\bar\nu_{l}$)
are produced. The states of 
flavor neutrinos and antineutrinos are {\em coherent superposition} of states of neutrinos with definite masses.
The probabilities  of the decay (\ref{9}) and other similar processes are given by the SM in which
$\nu_{l}$ and $\bar \nu_{l}$ can be considered as massless particles.

The condition under which the coherent neutrino flavor states 
(\ref{19}) are produced can be written in the form (see \cite{BilPon,Carlo})
\be
L_{0}\gtrsim d
\label{21}
\ee
where $d$ is a quantum-mechanical size of the source and 
\be 
L_{0}=4\pi \frac{E}{\Delta m^{2}}
\label{22}
\ee
is oscillation length {\em in the rest frame of the neutrino source}. Let us notice 
that in the case of the 
reactor KamLAND
experiment $L_{0}\simeq 10^{2}$ km and condition (\ref{21}) is obviously fulfilled.

\section{Neutrino oscillations}
Let us consider the evolution of a flavor neutrino state in vacuum. The evolution equation in the quantum field theory is the Schrodinger equation
\be
i\,\frac{\partial |\Psi(t) \rangle}{\partial t} = H\, |\Psi(t) \rangle,
\label{23}
\ee
where $ H$ is the total Hamiltonian. In the case of the vacuum  $ H= H_{0}$ and the general solution of the equation (\ref{23}) has the form 
\be
|\Psi(t) \rangle = e^{-i\,H_{0}\, t }\, |\Psi(0) \rangle ,
\label{24}
\ee
where $|\Psi(0) \rangle$ is the initial state. In our case 
\be
|\Psi(0) \rangle=|\nu_{l}\rangle, 
\label{25}
\ee
and the state $|\nu_{l}\rangle$ is given by the relation  (\ref{19}). Taking into account that 
\be
H_{0}\, ~|i \rangle =E_{i}\,~|i \rangle;~~E_{i}\simeq p +\frac{m^{2}_{i}}{2p}
\label{26}
\ee
for the state of neutrino at the time $t\geq 0$ we find
\be
|\nu_{l}\rangle_{t}
= \sum_{i=1}^{3}|\nu_i\rangle\, e^{-iE_{i}\,t}\,U_{l i}^* ,
\label{27}
\ee
Thus,  if at $t=0$ flavor neutrino is produced at time $t>0$ the state of neutrino is a 
{ \em superposition 
of stationary states}. Phases of different states of neutrinos with definite masses  at the time $t$ are different. This is  the main 
quantum mechanical reason for neutrino oscillations. 

Neutrinos are detected via the observation of CC and NC weak processes. Let us 
consider,  for example, the CC inclusive process
\be
\nu_{i}+N \to l' + X.
\label{28}
\ee
Because effect of small neutrino masses can not be detected,  for the matrix element of the process 
we have 
\be
\langle l'\,X\,|~ S~ |\,\nu_{i}\,N \rangle=U_{l'i}\, \langle l'\,X\,|~
S~ |\,\nu_{l'}\,N \rangle_{SM},\label{29}
\ee
where 
\be
\langle l'\,X\,|~ S~ |\,\nu_{l'}\,N\,\rangle_{SM}= 
-i\,\frac{G_{F}}{\sqrt{2}}\,N\,2\,
\bar u_{L}(p')\,\gamma_{\alpha}\,u_{L}(p)\,\langle X|\, J^{\alpha}(0)\,|N \rangle \,(2\pi)^{4}\,
\delta (P'-P).
\label{30}
\ee
is the SM matrix element of the process ($p'$ is the momentum of final lepton, 
$p$ is the momentum of neutrino).

From (\ref{27}) and (\ref{29}) for the probability of 
$\nu_{l} \to \nu_{l'}$ transition in vacuum we obtain the following expression
\be
{\mathrm P}(\nu_{l} \to \nu_{l'}) =|\,\sum^{3}_{i=1}U_{l'  i} \,~
e^{- i\,E_{i}t} \,~U_{l i}^*\, |^2.
\label{31}
\ee
Taking into account the unitarity of the mixing matrix we can easily see that 
probabilities, given by 
(\ref{31}),  are normalized: 
\be
\sum_{l'}{\mathrm P}(\nu_{l} \to \nu_{l'})= \sum_{l}{\mathrm P}(\nu_{l} \to \nu_{l'})=1.
\label{32}
\ee
Let us stress that probability of the decay in which flavor neutrinos  $\nu_{l}$ are produced and cross section of neutrino-absorption process in which leptons $l'$ are produced
are given by the standard model. 

From  (\ref{31}) we obtain the following standard expression for probability of the 
$\nu_{l} \to \nu_{l'}$ transition in vacuum
\be
P(\nu_{l} \to\nu_{l'}) =
|\delta_{l' l}+ \sum_{i=2,3} U_{l' i} \,(e^{\frac{ \Delta m^2_{ 1i }}{ 2 E }\, L }-1)
U^{*}_{li}|^{2},
\label{33}
\ee
where $L$
is the distance between neutrino-production and neutrino-detection points.\footnote{ We assume that $m_{1}< m_{2}<m_{3}$.}
\newpage
We took into account that \footnote{
This relation was used  (and tested) in the K2K 
experiment \cite{K2K}. 
In order to produce neutrino beam protons 
were extracted from KEK accelerator
in  1.1 $\mu sec$ spills 
every 2.2 $sec$.
Neutrino events which satisfy the criteria $-0.2\leq\Delta t \leq 1.3 \, \mu sec$  were selected in the 
experiment. Here
$\Delta t = t_{SK}-t_{KEK}-t_{TOF}$, $t_{KEK}$ is the measured time of the production of neutrinos at KEK, 
$t_{SK}$ is the  measured time of  the detection of neutrinos in the Super-Kamiokande detector and
$t_{TOF}=L/c\simeq 0.83\cdot 10^{3}\,\mu sec$
 is the time which required for neutrinos produced at the KEK to reach the SK. }
\be
L\simeq t
\label{34}
\ee
Analogously, for the transition $\bar\nu_{l} \to \bar\nu_{l'}$ we find
\be
P(\bar\nu_{l} \to\bar\nu_{l'}) =
|\delta_{l' l}+ \sum_{i=2,3} U^{*}_{l' i} \,(e^{\frac{ \Delta m^2_{ 1i }}{ 2 E }\, L }-1)
U_{li}|^{2},
\label{35}
\ee
From (\ref{33}) and (\ref{35}) it follows that  neutrino oscillations 
can be observed if (at least for one $i$) the condition 
\be
 \frac{ \Delta m^2_{ 1i }}{ 2 E }~ L  \gtrsim 1
\label{36}
\ee
is satisfied.

Taking into account  (\ref{1}) and (\ref{2}) we can conclude from  
(\ref{33}) and (\ref{35})
that in the atmospheric and accelerator long baseline region 
($ \frac{ \Delta m^2_{ 23 }}{ 2 E }~ L  \gtrsim 1$) leading transitions are 
$\nu_{\mu} \to\nu_{\tau}$
and $\bar\nu_{\mu} \to\bar \nu_{\tau}$
and $\nu_{\mu}$ ($\bar\nu_{\mu}$) survival probability  is given by
\be
{\mathrm P}(\nu_\mu \to \nu_\mu) = 
{\mathrm P}(\bar\nu_\mu \to \bar\nu_\mu)=
1 - \frac {1}
{2}\,\sin^{2}2\theta_{23}\, (1-\cos \Delta m_{23}^{2}\, \frac {L}
{2E}).\label{37}
\ee
In the reactor KamLAND region ($ \frac{ \Delta m^2_{ 12 }}{ 2 E }~ L  \gtrsim 1$) 
dominant transitions are 
$\bar\nu_{e} \to\bar \nu_{\mu}$ and 
$\bar\nu_{e} \to\bar \nu_{\tau}$
and for $\bar\nu_{e} $ survival probability
we have
\be
{\mathrm P}(\bar \nu_e \to \bar\nu_e)
=1-\frac{1}{2}~\sin^{2}2\,\theta_{12}~ (1 - \cos \Delta m_{12}^{2}
\,\frac {L}{2E}),\label{38} 
\ee
For solar neutrinos $\nu_{e} $ survival probability in matter in the leading approximation 
 is given by the two-neutrino 
expression which depend on $\Delta m_{12}^{2}$ and $\tan^{2}\,\theta_{12}$.
Existing neutrino oscillation data are well described by the leading approximation.

\section{Time-energy uncertainty relation. Translations}
There are two types of uncertainty relations in Quantum theory (see, for example, \cite{Sakurai}).
The Heisenberg uncertainty relations 
connect uncertainties of two measurable quantities with modulus of average value of their commutators.
If a and b are two non commuting hermitian operators, corresponding to two measurable quantities,
uncertainty of these quantities in any state are connected by the relation;
\be
\Delta a~\Delta b \geq \frac{1}{2}~|\overline{[a,b]}|
\label{39} 
\ee
Due to Heisenberg uncertainty relation 
$\Delta p~\Delta q \geq \frac{1}{2}$ it is impossible to distinguish production of neutrinos with different masses. As we discussed before, this is the reason why coherent flavor neutrino (antineutrino) states $|\nu_{l}\rangle$ ($|\bar\nu_{l}\rangle$) are produced.

For non stationary states with finite time interval $\Delta t$, during which
significant changes in the
 system happen,  time-energy relation
\be
\Delta E~\Delta t \geq 1
\label{40} 
\ee
takes place. Thus, finite time interval requires  uncertainty of energy.

Time-energy uncertainty relation and Heisenberg uncertainty relation have completely different origin. In quantum theory time is parameter and there is no operator 
which corresponds to time.  
Time-energy uncertainty relation  is valid for systems which are described by 
non stationary states.

If at $t=0$ flavor neutrino is produced,  at the time $t$
neutrino state will be  a superposition of stationary states which is given by Eq. (\ref{27}). 
From this equation it follows that flavor 
content of neutrino state can be changed significantly when time t satisfies the 
condition 
\be
\Delta E_{1i}~t \geq 1
\label{41} 
\ee
where $E_{i}-E_{1}\simeq \frac{\Delta m^{2}_{1i}}{2E}$.\footnote{It is obvious that (\ref{41}) is only necessary condition of the change of the flavor content of the state. It is also necessary that two corresponding elements of the neutrino mixing matrix are not small.} This condition  coincides
with inequality (\ref{36}), which is the condition to observe neutrino oscillations.

Let us stress again that for the time $t$, during which a significant change of the 
flavor  content of the state happen, to be finite, 
uncertainty of energy is needed. 

Conservation of energy and momentum is consequence of the invariance under translations. We will demonstrate now that in the case of the mixed flavor neutrino states 
there is no invariance under translations.

Let us consider translations of coordinates (see, for example \cite{BogShir})
\be
x'_{\alpha} = x_{\alpha} +a_{\alpha},
\label{42}
\ee
where $a$ is a constant vector. If invariance under translations holds  
we have
\be
|\Psi \rangle' = e^{i\,P\, a }\, |\Psi \rangle,
\label{43}
\ee
where vectors $|\Psi \rangle $ and $|\Psi \rangle'$ describe the {\em the same} physical states and 
$P_{\alpha}$ is the operator of the total momentum. If $|\Psi \rangle $ is the state with total momentum
$p$ vectors $|\Psi \rangle'$ and $|\Psi \rangle'$ differ by the phase factor
\be
|\Psi \rangle' = e^{i\,p\, a }\, |\Psi \rangle .
\label{44}
\ee
In the case of the invariance under translations for any operator $O(x)$ we have
\be
O(x+a)=  e^{i\,P\, a }\,O(x)\, e^{-i\,P\, a }
\label{45}
\ee
From this relation it follows
\be
i~\partial_{\alpha}O(x)= [O(x), P_{\alpha}]
\label{46}
\ee
and 
\be
O(x)=  e^{i\,P\, x }\,O(0)\, e^{-i\,P\, x }.
\label{47}
\ee
Let us apply now the operator of the translations $e^{i\,P\,a}$ to the mixed flavor neutrino state 
$|\nu_{l}\rangle$. 
We have
\be
|\nu_{l}\rangle'=    e^{i\,P\, a }\, |\nu_{l}\rangle =  e^{-i\,\vec{p}\,\vec{a}}\,\sum_{l'} 
|\nu_{l'}\rangle \,\sum_{i}U_{l' i}e^{i\,E_{i}\,a}\,U_{l i}^*,
\label{48}
\ee
i.e. the  vector $|\nu_{l}\rangle'$ describes a superposition of different flavor states. Thus 
initial and transformed vectors describe {\em different states}.
We come to the conclusion that in the case of the mixed flavor states
there is no invariance under translations. This 
means that in   transitions between different flavor neutrinos energy is not conserved.
Non conservation of energy in neutrino oscillations is obviously connected with finite time between
neutrino production and neutrino detection and with the time-energy 
uncertainty relation (\ref{41}). 

We would like to finish with the following remark. In several papers (see \cite{Kaiser})
 it is claimed that in neutrino oscillations the energies of neutrinos with different masses must be the same. It is obvious that this is impossible 
from the point of view of time-energy uncertainty relation discussed here. 

We will 
present another argument against equal energies assumption. Let us consider usual Hamiltonian of neutrino in matter in the flavor representation. It is the sum of two terms: free Hamiltonian and interaction Hamiltonian. For free Hamiltonian we have
\be 
\langle \nu_{l'}|H_{0}|\nu_{l}\rangle = \sum_{i}
U_{l'i}~E_{i}~U^{*}_{li}
\label{49} \ee 
If  energies of neutrinos with definite masses are the same ($E_{i}=E$), the free Hamiltonian is unit matrix
\be 
\langle \nu_{l'}|H_{0}|\nu_{l}\rangle =
E~\delta_{l'l}
\label{50} \ee 
In this case it would be no MSW matter effect \cite{MSW} observed in solar neutrino experiments 
\cite{Lisi}.

\section{Conclusion}

Neutrino oscillations were observed in SK, SNO, KamLAND and other neutrino experiments.
It  was proved by these experiments that two neutrino  mass-squared differences are different from zero.
This means that neutrinos are particles with different from zero masses. At present we do not know the values of neutrino masses. However, from the data of tritium experiments and cosmological data we know that neutrino masses must be 
smaller than $\simeq 1$ eV. Thus, it was discovered that  neutrino masses are nonzero and 
 many orders of magnitude smaller than masses of leptons and quarks.
It is a general opinion that the  explanation of this smallness requires a new beyond the SM scale.

Basics of neutrino oscillations are still under active discussions. 
From the point of view considered here neutrino oscillations are based on 
two uncertainty relations: Heisenberg uncertainty relation and time-energy uncertainty relation.
Flavor neutrinos $\nu_{e}$, $\nu_{\mu}$ and $\nu_{\tau}$
are produced in CC weak processes together with, correspondingly, $e^{+}$, $\mu^{+}$ and  
$\tau^{+}$. Because of the Heisenberg uncertainty relation  
the states of the flavor neutrinos are coherent superposition of states of neutrinos with definite masses.

Phenomenon of neutrino oscillations with finite oscillation time (length) is  due to 
time-energy uncertainty relation which requires energy uncertainty. 
Time-energy uncertainty relation is based on the fact that state of neutrino, produced (at $t=0$)
as a flavor state, at the time $t$ is {\em a superposition of the stationary 
states}. 

I am thankful to S.T. Petcov for numerous discussions.
I would like to acknowledge 
the Italian program
``Rientro dei cervelli'' for the support.

\end{document}